\documentclass{sf2a-conf2015}
\usepackage{graphicx}
\usepackage{hyperref}
\usepackage[]{natbib}  
\usepackage{epstopdf}
\usepackage{caption}
\def\BibTeX{{\rm B\kern-.05em{\sc i\kern-.025em b}\kern-.08em
    T\kern-.1667em\lower.7ex\hbox{E}\kern-.125emX}}
\bibpunct{(}{)}{;}{a}{}{,}  
\newcommand{\kms}{{\mathrm{km~s^{-1}}}}

\newcommand{\angstrom}{\mbox{\normalfont\AA}}

\begin{document}

\TitreGlobal{SF2A 2015}


\title{Estimating Stellar Fundamental Parameters Using PCA: Application to Early Type Stars of GES Data}

\runningtitle{PCA-based inversion of stellar parameters}

\author{Farah W.}\address{Department of Physics \& Astronomy, Notre Dame University - Louaize, Lebanon}

\author{Gebran M.$^{1,}$}



\author{Paletou F.}\address{Universit\'e Paul Sabatier, OMP \& IRAP, CNRS, F--31400 Toulouse}
\author{Blomme R.}\address{Royal Observatory of Belgium, Ringlaan 3, B-1180 Brussels, Belgium}

\setcounter{page}{237}


\maketitle

\begin{abstract}
This work addresses a procedure to estimate fundamental stellar parameters such as $T_{\rm eff}$, log\textit{g}, [Fe/H], and $v\sin i$ using a dimensionality reduction technique called principal component analysis (PCA), applied to a large database of synthetic spectra. This technique shows promising results for inverting stellar parameters of observed targets from Gaia Eso Survey.
\end{abstract}



\section{Introduction}
With the introduction of new telescopes and instruments to the scientific astronomical community, and the rapid increase of sky surveys such as SDSS and RAVE, tremendous amount of spectral data is being acquired on a daily basis and with an increasing rate. 
Therefore, these challenges urged the need of efficient and automated techniques to handle and analyze this huge amount of information. Such automated procedures for classification of stars have been discussed recently using different codes and mathematical approaches. As an example, one can mention the methods used to analyze a spectral library described in \cite{Jofre2014}.\\
In this work, we present a dimensionality reduction technique called PCA, applied to a huge database of synthetic spectra. PCA searches in a high dimensional space for possible correlations, and finds an optimal basis for representing the data in a compact way.
Due to the high number of spectra in each synthetic database ($\sim$200\,000), and the high number of data points in each spectral domain ($\sim$2\,500. Same as observation, see section 2), such technique is crucial for inverting stellar parameters of observed targets from Gaia ESO survey. Using PCA, data can be represented in a fewer number of data points, allowing a fast ``nearest neighbor(s)" search between the observed data set and the synthetic spectra. This study is an extension of \cite{Paletou2015} where the H-R domain of application has been extrapolated to stars of types earlier than F, and the training database used in this work is a set of synthetic spectra.

\section{Observation}

The procedure is applied to more than 800 stars, members of the open clusters NGC3293, NGC6705, and Trumpler14. The observations are part of the GAIA ESO public survey and consist of 2 spectral ranges, one samples the H$_{\delta}$ line region [4030-4200]\angstrom and the second samples the [4400-4550]\angstrom (HR5) region. These spectra were taken using GIRAFFE/FLAMES spectrograph at a resolution R $\approx$ 25\,000, and reduced by the GES.

\section{Spectral range selection}

Balmer lines, due to the broadening caused by the Stark's effect, are excellent indicators of effective temperature and surface gravity \citep{Graybook}. The reason behind studying H$_{\delta}$ in particular is because this line is formed in deep enough atmospheric layers where LTE can still be considered as a reasonable assumption. Moreover, the HR5 region was chosen since metallic lines (namely Fe\,{\sc ii}, Mg\,{\sc ii}, Ti\,{\sc ii}, ...) are potentially good indicators of rotational velocity and metallicity.

\section{Synthetic Spetra}
LTE model atmospheres were calculated using ATLAS9 code \citep{Kurucz1992} and were used as input to the spectrum synthesis code SYNSPEC48 \citep{Hubeny1992} in order to compute a large grid of synthetic line profiles, over the same spectral regions as the observations. Spectra were calculated for T$_{eff}$ between 5\,000 and 15\,000 K, gravities between 2.0 and 5.0 cgs, rotational velocities between 0 and 200 $\kms$, and metalicities between -0.6 and 0.4 dex (only for the HR5 region, whereas a solar [Fe/H] was assumed for the $H_{\delta}$ region), all at a microturbulence of 2 $\kms$ and at a resolution of 25\,000.

\section{Procedure}
The central idea of principal component analysis is to reduce the dimensionality of a data set in which there are a large number of interrelated variables, while retaining as much as possible of the variation present in the data set \citep{PCAbook}. PCA searches for basis vectors that represent most of the variance in a given database. These vectors ($\vec{e}_k$) are in fact the eigenvectors of the variance-covariance matrix (\ref{var-covar}) of the synthetic data set \textbf{S}.
\begin{equation}
C=(\textbf{S}-\bar{\textbf{S}})^T.(\textbf{S}-\bar{\textbf{S}})
\label{var-covar}
\end{equation}
Where $\bar{\textbf{S}}$ being the mean spectrum over all the database.\\
Once the basis is obtained (adopted a set of 12 vectors, i.e $k=1,..,12$), the synthetic spectra and each observation (O) are projected unto this basis to obtain the projected coefficients (\ref{proj synth} \& \ref{proj obs})
\begin{equation}
p_{j,k}=(\textbf{S}_j-\bar{\textbf{S}}).\vec{e}_k \hspace{0.2cm} \footnote{$\textbf{S}_j$ is the $j^{th}$ spectrum (a row vector) in the database (matrix) \textbf{S}.}
\label{proj synth}
\end{equation}

\begin{equation}
\rho_{k}=(\textbf{O}-\bar{\textbf{S}}).\vec{e}_k
\label{proj obs}
\end{equation}
Then, a standard chi-squared (\ref{chi-sqr}) is performed in this low dimensional space in order to achieve a fast inversion of stellar parameters of the observed targets. The parameters of the synthetic spectrum having the minimum \textit{d} will be considered as the observation fundamental parameters.
\begin{equation}
d_j^{(O)}=\Sigma_{k=1}^{12}(\rho_k-p_{j,k})^2
\label{chi-sqr}
\end{equation}
The observation spectra were radial velocity corrected, and those with low signal-to-noise ratio were filtered out. Upon starting the inversion process, the technique showed to be very sensitive to normalization of spectra, thus an iterative ``re-"normalization procedure was performed according to \cite{Gazzano2010}.

\section{Results}
\begin{figure}[]
 \centering
 \includegraphics[width=0.8\textwidth,clip]{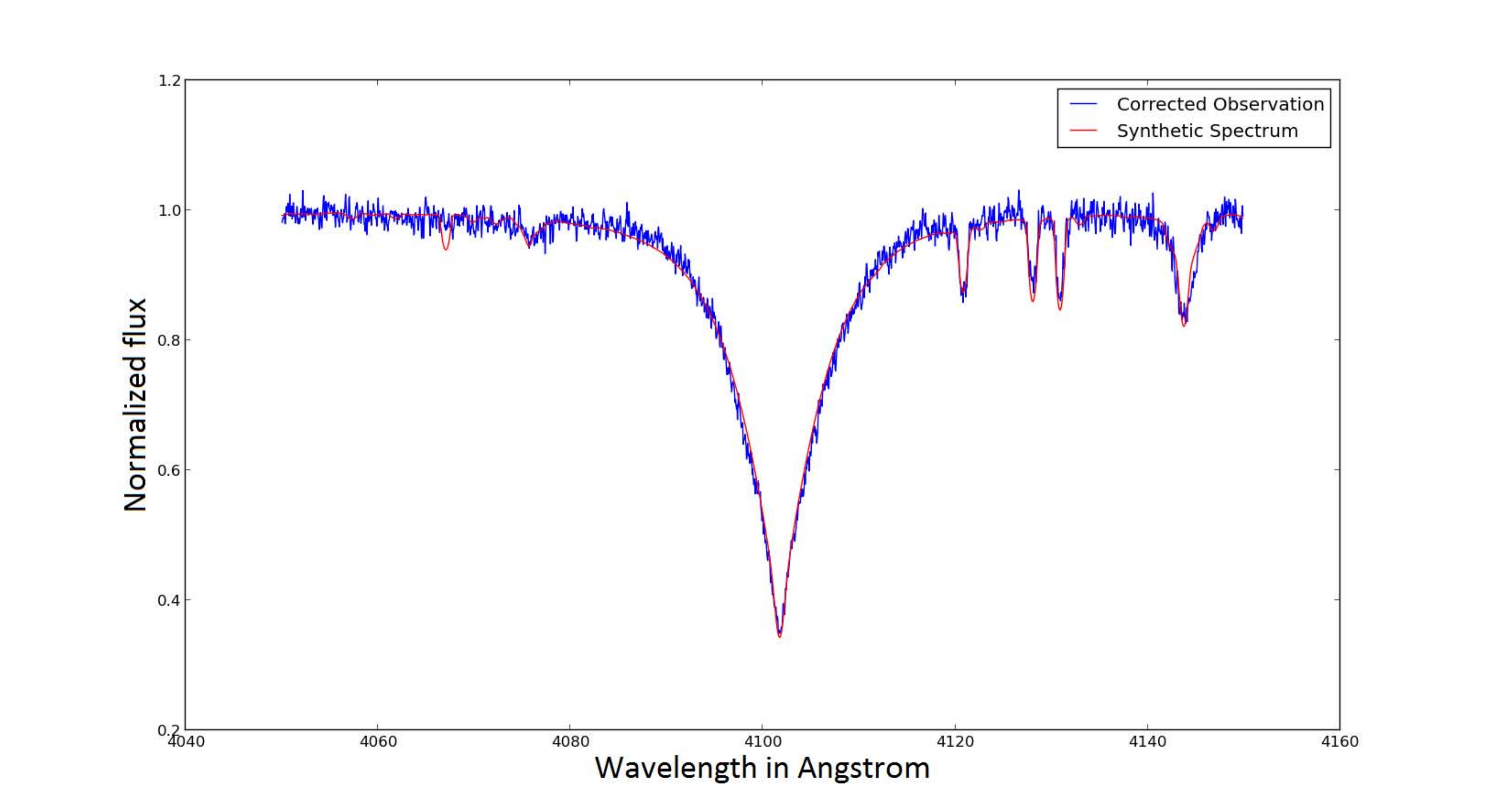}      
  \caption{Example of the fitting of the H$_\delta$ line of the star 10361733-5809031 member of NGC3293 cluster, with a synthetic spectrum. Blue being the observed spectrum, while red the fitted synthetic.}
  \label{Hdelta}
\end{figure}

\begin{figure}[]
 \centering
 \includegraphics[width=0.8\textwidth,clip]{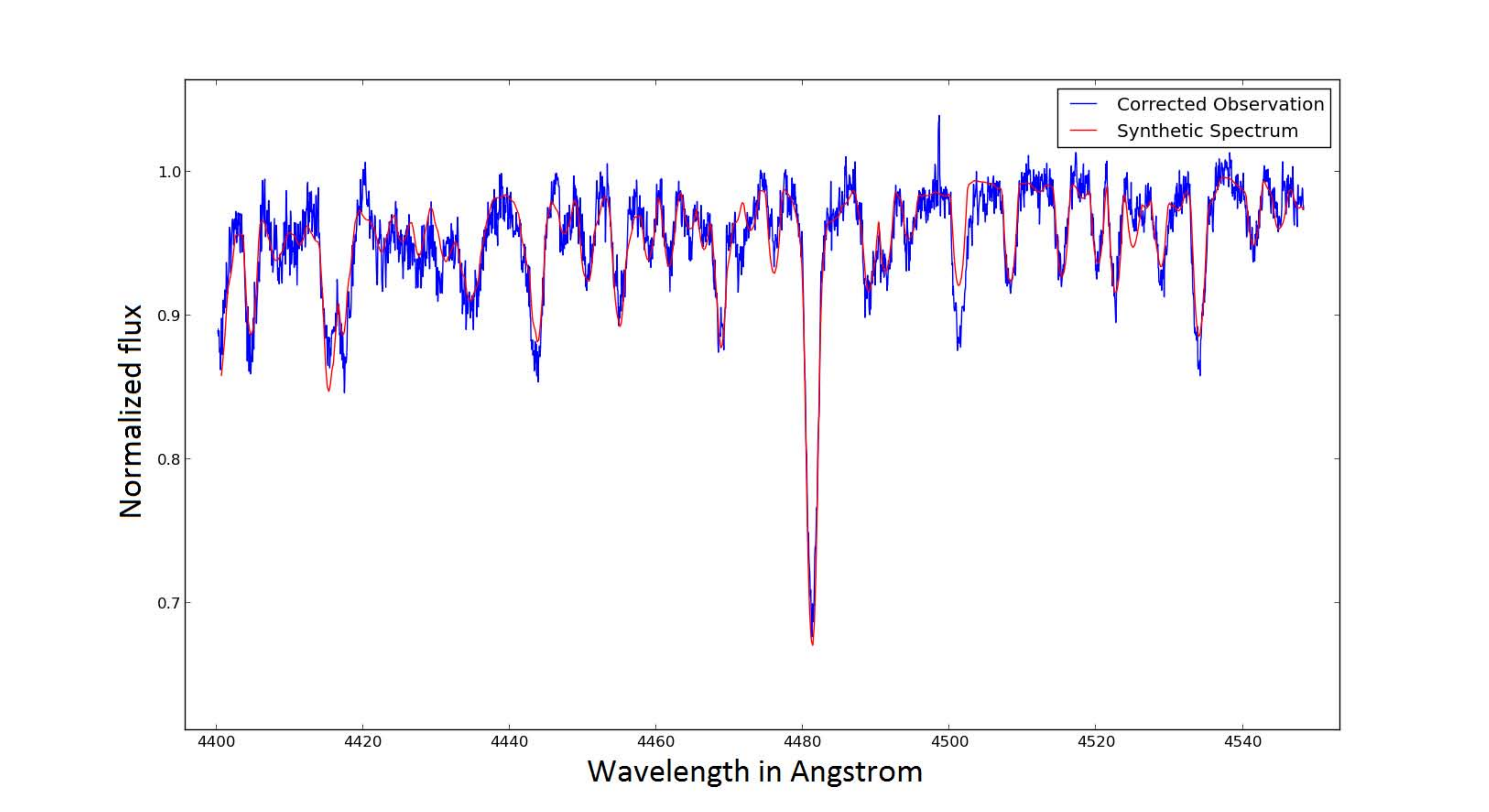}      
  \caption{Example of the fitting of the region containing Fe\,{\sc ii}, Mg\,{\sc ii} and Ti\,{\sc ii} lines of the star 10430337-5941536 member of Trumpler14 cluster, with a synthetic spectrum. Blue being the observed spectrum, while red the fitted synthetic.}
  \label{met}
\end{figure}

In general, inversion based on this technique was performed over the selected stars, and the fundamental parameters of the targets were estimated. An example of the nearest neighbor search is given in figures \ref{Hdelta} and \ref{met}. The parameters derived by PCA, along with the non-official parameters obtained by WG13 of GES are detailed in table \ref{table}.
\begin{table}[h]
\centering
\caption{Results derived using PCA along with parameters given by WG13 of GES}

\begin{tabular}{|l|cccc|cccc|}
\hline

& \multicolumn{4}{c|}{Derived} & \multicolumn{4}{c|}{Given}  \\
\hline
Star & $T_{\rm eff}$ & log\textit{g} & $v\sin i$ & [Fe/H] & $T_{\rm eff}$ & log\textit{g} & $v\sin i$ & [Fe/H] \\
& (K) & (dex) & ($\kms$) & (dex) & (K) & (dex) & ($\kms$) & (dex) \\

\hline
10361733-5809031 & 14\,400 & 3.6 & 45 & - & 14\,775 & 3.84 & 44 & - \\
 
10430337-5941536 & 9\,200 & 4.6 & 70 & 0.4 & 8\,633 & 3.5 & 75 & - \\
\hline
\end{tabular}
\label{table}

\end{table}

\noindent With PCA, we will be contributing by determining stellar parameters to the next GES data release.



\section{Conclusions and future work}

PCA proved to be a fast and reliable inversion technique, with an ease to implement. An attempt to increase the size of the synthetic database is being performed in order to improve the accuracy in the parameters obtained. Moreover, the merging of two spectral ranges in a one data set is also considered as a future work.

\begin{acknowledgements}
This work is based on observations collected with the FLAMES spectrograph at the VLT/UT2 telescope (Paranal Observatory, ESO, Chile), for the Gaia-ESO Large Public Survey, programme 188.B-3002.
\end{acknowledgements}

\newpage

\bibliographystyle{aa}  

%
\end{document}